\documentstyle[epsf,12pt]{article}
\newcommand {\beq}{\begin{equation}}
\newcommand {\eeq}{\end{equation}}
\newcommand {\beqa}{\begin{eqnarray}}
\newcommand {\eeqa}{\end{eqnarray}}
\newcommand {\n}{\nonumber \\}

\renewcommand{\theequation}{\thesection.\arabic{equation}}
\begin{document}
\setlength{\oddsidemargin}{0cm}
\setlength{\baselineskip}{7mm}

\begin{titlepage}
 \renewcommand{\thefootnote}{\fnsymbol{footnote}}
$\mbox{ }$
\begin{flushright}
\begin{tabular}{l}
KEK-TH-1048\\
NTNU-phy-05-10-01\\
Oct.  2005
\end{tabular}
\end{flushright}

~~\\
~~\\
~~\\

\vspace*{0cm}
    \begin{Large}
       \vspace{2cm}
       \begin{center}
{Fuzzy Spacetime with $SU(3)$ Isometry\\ in IIB Matrix Model}
\\
       \end{center}
    \end{Large}

  \vspace{1cm}

\begin{center}
           Hiromichi K{\sc aneko}$^{2)}$\footnote
           {
e-mail address : kanekoh@post.kek.jp},
            Yoshihisa K{\sc itazawa}$^{1),2)}$\footnote
           {
e-mail address : kitazawa@post.kek.jp}\\{\sc and}
           Dan T{\sc omino}$^{3)}$\footnote
           {
e-mail address : dan@home.phy.ntnu.edu.tw}

        $^{1)}$ {\it High Energy Accelerator Research Organization (KEK),}\\
               {\it Tsukuba, Ibaraki 305-0801, Japan} \\
        $^{2)}$ {\it Department of Particle and Nuclear Physics,}\\
                {\it The Graduate University for Advanced Studies,}\\
                {\it Tsukuba, Ibaraki 305-0801, Japan}\\
                 
        $^{3)}$ {\it Department of Physics, National Taiwan Normal University,}\\
                {\it Taipei 116, Taiwan }

\end{center}

\vfill

\begin{abstract}
\noindent A group of fuzzy spacetime with $SU(3)$ isometry
is studied at the two loop level in IIB matrix model. 
It consists of spacetime from 4 to 6 dimensions, namely from $CP^2$
to $SU(3)/U(1)\times U(1)$. The
effective action scales in a universal manner in the large $N$ limit
as $N$ and $N^{4/3}$ on 4 and 6 dimensional manifolds respectively. 
The 4 dimensional spacetime
$CP^2$ possesses the smallest effective action in this class.
\end{abstract}
\vfill
\end{titlepage}
\vfil\eject

\section{Introduction}
\setcounter{equation}{0}
\setcounter{footnote}{0}

The investigations of the properties of the spacetime at the
microscopic level have become an important physical subject since
we now have a clear picture where the universe comes from and is going.
At the current stage, the space is found to be almost flat and accelerating its
expansion rate. It is therefore approaching a 4 dimensional de Sitter
spacetime. Furthermore the scale independent fluctuation of the
cosmic microwave background at long distance scale suggests that the universe
also started as a de Sitter spacetime. In order to explain why the universe
evolves in such a peculiar way, we need to obtain a deeper understanding of the spacetime.
It is expected that string
theory plays a crucial role to understand the spacetime at the
microscopic level. In order to address a time dependent issue, it is 
likely that we need a non-perturbative formulation of string theory such as
IIB matrix model\cite{IKKT96,FKKT98}.

In this model, Euclidean spacetime is expected to emerge out of
the distributions of the
eigenvalues of the 10 matrices.
We can certainly imagine that the eigenvalues are homogeneously distributed
on $S^4$ in 10 dimensions.
Since a de Sitter space becomes an $S^4$ after the Euclidean continuation,
we may interpret Euclidean spacetime a la Hartle and Hawking\cite{HW}.
If we divide an $S^4$ into the two halves, we obtain an $S^3$ at the boundary.
With the identification of the $S^3$ as a space, the effective action for $S^4$
in IIB matrix model determines the relative probability of the
emergence of an $S^3$ out of nothing.
We find it remarkable that the matrix models can accommodate
a realistic spacetime in a nonperturbative way.
In this sense our studies of homogeneous spacetime in IIB matrix model
may shed light on the origin of the universe.

A fuzzy homogeneous spacetime ${G}/{H}$ can be embedded in matrix
models by choosing background matrices as 
the generators of a group $G$ \cite{matrix
homogeneous}. $G$ has to be a subgroup of $SO(10)$ and $H$ has to be
a closed subgroup of $G$. We obtain Non-Commutative (NC) gauge
theory on the fuzzy spacetime in this construction \cite{NCgauge}.
We can calculate an effective action on this background and investigate
the large $N$ scaling behavior of it.

In this paper, we choose $G$ to be $SU(3)$ and investigate the class
of the manifolds with $SU(3)$ isometry in IIB matrix model. They
include $CP^2={SU(3)}/{U(2)}$ and ${SU(3)}/{U(1) \times U(1)}$. Each
manifold is labeled by an irreducible representation of $SU(3)$.
Note that $CP^2$ is a four dimensional manifold, while ${SU(3)}/{U(1)
\times U(1)}$ is six dimensional. Therefore we can investigate the
large $N$ scaling behavior of the effective action for the both 4
and 6 dimensional manifolds.

In a series of papers \cite{fuzzy sphere}, we investigated
the manifolds with $SU(2)\times SU(2)$ isometry and found certain
instabilities associated with fuzzy $S^2\times S^2$. Each fuzzy
$S^2$ can be parameterized by $l$: the spin of a representation
and $f$: a scale factor. We recall that the radius of $S^2$ is $lf$
while the NC length scale is $\sqrt{lf}$. Thus the both spin and
scale factor specify the overall size of each $S^2$. In this construction
$S^2 \times S^2$ can be characterized by the ratios of the spins and
scale factors between the two $S^2$'s. The instability has been
found under the variation of the both ratios. However it 
does not take place if we are constrained to have the
identical scale factor for the both $S^2$'s. We thus expect that a
more symmetric manifold will be stable.

In this respect $CP^2$ backgrounds are interesting. $CP^2$ can be
embedded in Hermitian matrices as \beq A_i =f T_i ,\eeq where $T_i$
are the generators of $SU(3)$ in a particular class of representations. 
As $CP^2$ can have the only one scale
factor, it may not suffer from such an instability. 
The irreducible representations of $SU(3)$ from which $CP^2$ can be
constructed as $SU(3)/U(2)$ are relatively well
studied\cite{ABNNcp2,SU(3)}. Therefore it is interesting to
investigate the large $N$ scaling behavior of the effective action
of $CP^2$ and other manifolds with $SU(3)$ isometry and to see which
manifold is most stable among them.

The organization of this paper is as follows. In section
\ref{MTonCP}, we construct IIB matrix model on fuzzy $CP^2$. We find
a universal expression for the 2-loop effective action on a
homogeneous space. In section \ref{EFonCP}, we derive the effective
actions on the manifolds with $SU(3)$ isometry and investigate the
large $N$ scaling behavior of them. 
We find that they scale in a universal fashion which depends only
on the dimensionality of the manifold.
We argue that there is indeed a universality in the large $N$
scaling of the effective action on $G/H$.
We conclude in section \ref{conc} with discussions. 
In Appendix A, we construct the generators and
eigen-matrices of $SU(3)$. In Appendix B, we derive the 2-loop
effective action on the manifolds constructed from $SU(3)$ algebra. In
Appendix C, we numerically evaluate the 2-loop effective action.

\section{I{}IB matrix model on fuzzy $CP^2$} \label{MTonCP}
\setcounter{equation}{0}
\subsection{Group theoretic construction}
Let us recall a construction of fuzzy homogeneous spacetime $G/H$
and gauge field on them\cite{matrix homogeneous}. We pick a state
$|0\rangle$ in a definite representation $G$ which is invariant
under $H$. The set of all states which can be reached by multiplying
elements of $G$ to $|0\rangle$ is called the orbit of $|0\rangle$. A
fuzzy homogeneous spacetime $G/H$ is constructed as the orbit of
$|0\rangle$. It is represented by the irreducible representation
that is descended from $|0\rangle$. The basic degrees of freedom in
NC gauge theory are bi-local fields. We construct NC gauge field as
the bi-local field by forming the tensor product of the relevant
irreducible representation and its complex conjugate.

We take a Lie group $G$ to be $SU(3)$ in the present investigation.
An irreducible representation of $SU(3)$ is labeled by a set of two
integers $(p,q)$. An invariant subgroup $H$ depends on the
irreducible representation. We have $U(2)$ as the invariant subgroup
for a $(p,0)$ representation. It gives rise to a four dimensional
fuzzy $CP^2=SU(3)/U(2)$. 
On the other hand $H$ is $U(1)\times U(1)$
for a generic $(p,q)$ representation. In this case we obtain a fuzzy
flag manifold ${SU(3)}/{U(1)\times U(1)}$. It is a six dimensional
NC spacetime which locally looks like $CP^2\times S^2$. The
representation $(p,p)$ may give the most symmetric six dimensional
manifold.

In the large $N$ limit,
the extension of the manifold becomes infinite with respect to
NC scale. In such a situation,
we expect that the effective action scales in a definite way.
As we find such a scaling exhibits a universality which depends only on
the dimensionality of the manifold, a group of the representations
represents a universal class. We are thus interested in to identify
such a universal manifold in the large $N$ limit.

We introduce a fuzzy homogeneous spacetime as a background of I{}IB
matrix model, and calculate the effective action in a background
field method. For this purpose,
we expand the matrices around the background with a scale factor $f$:
\begin{eqnarray}
A_{\mu}=f(p_{\mu}+a_{\mu}) ,
\label{9.19.1}
\end{eqnarray}
where $p_{\mu}$ is the background and $a_{\mu}$ represents NC gauge
field. The background is taken as
\begin{eqnarray}
p_{\mu}&=&
\left\{
\begin{array}{cc}
{\bf 1}_{n\times n}\otimes T_{\mu}^{(p,q)} &\mu=1, ..., 8\\
0& \mu=9,10
\end{array}\right .
\label{9.19.2}
\end{eqnarray}
where $T_{\mu}^{(p,q)}$ are the $SU(3)$ generators of a $(p,q)$
representation. Here we have taken a simple reducible representation. 
We obtain $U(n)$ gauge theory on a fuzzy homogeneous
spacetime in this way. This background can be realized by the matrices
whose dimension is \beq N=n\cdot
dim(p,q)=n\cdot\frac{1}{2}(p+1)(q+1)(p+q+2). \eeq One could consider
a more general background such as
\begin{eqnarray}
\sum_{i}\oplus  \left({\bf 1}_{n_i\times n_i}\otimes T_{\mu}^{(p_i,q_i)}\right) \;.
\end{eqnarray}
However we consider a simple case (\ref{9.19.2}) only in the present paper.

The gauge field is expanded by harmonic functions on the $(p,q)$
background:
\begin{eqnarray}
a_{\mu}&=&\sum_{A} a_{\mu}^{(A)}Y^{(A)} ,
\label{9.19.3}
\end{eqnarray}
where the harmonic function matrices $Y^{(A)}$ are the eigen-functions
of $[T_3, \;]$, $[T_8,\;]$ and $[T_{\mu}, [T_{\mu},\;]]$. The
quantum numbers $(A)$ are determined by decomposing the gauge field
into the irreducible representations. An explicit construction procedure of
them is explained in appendix A. We obtain the propagators
and vertices by using the expansion (\ref{9.19.3}). By a
perturbative calculation, we obtain the effective action
$\Gamma=\Gamma(p,q,\lambda^2,n)\;.$ Here $\lambda^2$ is a natural
expansion parameter which is proportional to $1/f^4$. It is a 't
Hooft coupling constant which should be kept fixed in the large $N$
limit. We can determine the parameters $\{p,q,\lambda,n\}$ by requiring
that the effective action is stationary with respect to the change
of them $\delta\Gamma=0$. Such a set constitutes a solution of IIB
matrix model. Dynamical generation of fuzzy homogeneous spacetime
can be investigated in this way. We can compare the extremal values
of the effective action for these (stable) solutions to find the most favored one.

In this paper we carry out the loop expansion up to the 2-loop
level. The tree level action dose not admit a non-trivial solution.
Such a solution appears when the 2-loop quantum correction is
included in the effective action. The situation is the same with the
backgrounds based on $SU(2)$ algebras \cite{fuzzy sphere}
and, as we discuss later, a common aspect for backgrounds based on
Lie algebras $G\subset SO(10)$.

In what follows, we explain the details of our evaluation of the effective action.

\subsubsection*{Universal properties of the 2-loop effective action}
We can draw some common features of the effective action in
homogeneous spacetime from a series of our studies. Here we assume
the expansion (\ref{9.19.1}) and $p$ denotes a set of generators of
a Lie algebra $G \subset SO(10)$ of the form (\ref{9.19.2}). We also
assume that one can find a set of harmonic functions which are
eigenfunctions of the adjoint operators $P=[p, \;]$. In the large
$N$ limit, the leading terms of the effective action of I{}IB matrix
model up to the 2-loop level can be summarized as the following universal expression
\footnote{$G=SU(2)$ is the exception since the two loop amplitude is
finite in the large $N$ limit. We must use the exact propagators for
gauge bosons and fermions to evaluate the 2-loop contributions in
such a case. }
\begin{eqnarray} \label{ef_ac_gen}
\Gamma&=&\frac{f^4}{4}C_GC_2(G,R)N
+n^2\;O\left(tr \frac{1}{P^{4}}\right)
+2n^3\frac{C_G}{f^4}
\left\langle\frac{1}{P_1^2P_2^2P_3^2}\right\rangle
\;
\label{9.19.4}
\end{eqnarray}
where $R$ denotes an irreducible representation of a Lie algebra and
\begin{eqnarray}
C_G\;\delta_{\rho\sigma}= f_{\mu\nu\rho}f_{\mu\nu\sigma}
\;,\quad
C_2(G,R)N= tr\;p_{\mu}p^{\mu} .
\end{eqnarray}
$f_{\mu\nu\rho}$ is the structure constant of the Lie algebra. The
first, second and third terms in (\ref{9.19.4}) are the tree, 1-loop
and 2-loop contributions respectively.

The 2-loop contributions consist of the
planer and non-planar contributions. In NC theory, the non-planar
contributions are suppressed due to the NC phase. We
argue that the upper cut-off becomes $\sqrt{l}$ instead of $l$ in
the non-planar sector since the NC scale is $\sqrt{l}$. As the two
loop contributions are quadratically divergent in the large $N$
limit for a 4 dimensional background, we argue that the non-planar
contributions are suppressed by $\sqrt{N}$ in that case.  The
analogous suppressions should take place in higher dimensions. The
two loop non-planar contributions will be suppressed by $N$ in
comparison to the planar contributions for 6 dimensional backgrounds. We thus
argue that the 2-loop contributions are always positive since the
non-planar contributions can be neglected in the large $N$ limit.

The two loop level effective action can be bounded as
\begin{eqnarray} \label{ef_minac}
\Gamma&\ge& (\mbox{1-loop}) +2C_G\sqrt{
\frac{C_2(G,R)Nn^3}{2}
\left\langle\frac{1}{P_1^2P_2^2P_3^2}\right\rangle }\;.
\end{eqnarray}
after we minimize it with respect to $f$. Without the 2-loop
contributions, we can obtain only trivial solutions as $f=0$ is
required to minimize the action . Therefore higher loop,
at least 2-loop, corrections are necessary to obtain a fuzzy
homogeneous spacetime in I{}IB matrix model.

\section{The effective action on fuzzy spacetime with $SU(3)$ isometry.} \label{EFonCP}
\setcounter{equation}{0}

In this section, we evaluate the effective action on the fuzzy
manifolds with $SU(3)$ isometry. We set $n=1$ for simplicity since we can easily recover
the $n$ dependence as (\ref{ef_ac_gen}).

The tree level effective action of a $(p,q)$ representation is \beqa
\label{eff-tree} \Gamma_{tree} &=& - \frac{1}{4} Tr
[p_{\mu},p_{\nu}]^2 \n &=& \frac{3 f^4}{4} N \frac{1}{3} \left[ p
\left( p+3 \right) + q \left( q+3 \right) +pq \right] . \eeqa When
the background is $CP^2$ ($(p,0)$ rep.), the leading term of
(\ref{eff-tree}) in the large $N$ limit becomes \beqa \Gamma_{tree}
&\simeq& \frac{f^4}{2} N^2, \n N &\simeq& \frac{p^2}{2}. \eeqa On a
6d manifold ($(p,p)$ rep.), it becomes \beqa
 \Gamma_{tree} &\simeq& \frac{3f^4}{4} N^{\frac{5}{3}}, \n
 N &\simeq& p^3.
\eeqa

The leading term of the one loop effective action in the large $N$
limit can be estimate as
\beqa \Gamma_{1-loop} &\propto& Tr \left(
\frac{1}{P^2} \right)^2 \sim \left\{
\begin{array}{cc}
O \left( \log N \right) & \mathrm{CP^2} \n
O \left( N^{\frac{1}{3}} \right) & \mathrm{6d}
\end{array} \right. .
\eeqa
We can neglect this term in the effective action because
we shortly find that the effective action scales as $O(N)$ 
on $CP^2$ or $O(N^{4/3})$ on a 6
dimensional manifold.

The leading term of the two loop effective action in the large $N$
limit is evaluated as \beq \label{eff-2loop} \Gamma_{2-loop} =
\frac{6}{f^4} F_3\equiv\frac{6}{f^4}\left\langle{1\over
P_1^2P_2^2P_3^2}\right\rangle , \eeq where the detailed calculations
are explained in  appendix B. In this way, we obtain the effective
action in the large $N$ limit as \beqa \Gamma &=& \Gamma_{tree} +
\Gamma_{2-loop} \n &=& \frac{f^4N}{4} \left[ p \left( p+3 \right) +
q \left( q+3 \right) +pq \right] + \frac{6}{f^4} F_3. \eeqa

\begin{figure}[tbph]
\epsfysize=8cm
\begin{center}
\vspace{1cm}
\hspace{0cm}
\epsfbox{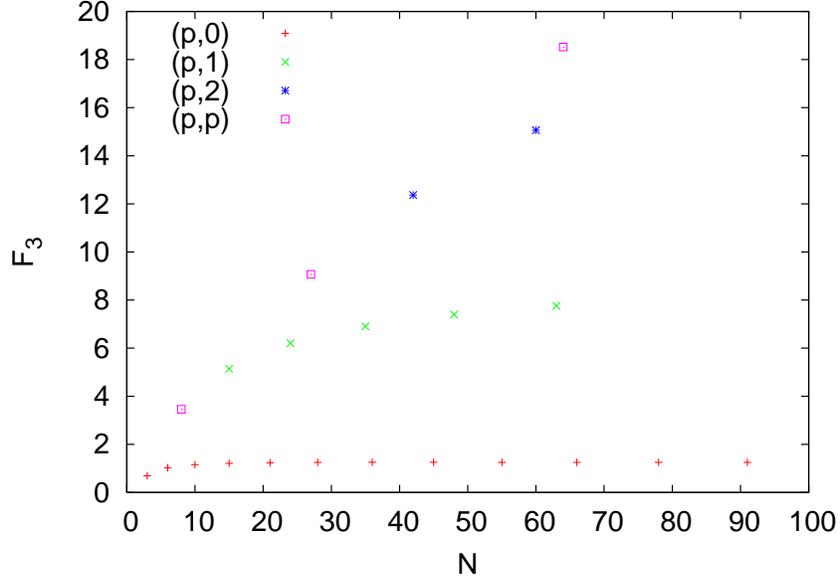}
\end{center}
\caption{$F_3$ against $N$.}
\label{fig:F_3}
\end{figure}

We can now explore the behavior of the effective action. Firstly, we investigate $F_3$ of
(\ref{eff-2loop})
 to determine the scaling behavior for various representations.
 We have numerically estimated
 $F_3$  in appendix C.
 Fig.\ref{fig:F_3} shows $F_3$ against $N$. We first
 observe that $F_3$ of the $(p,0)$ representations approaches a constant
 in the large $N$ limit.
This value is estimated as \beq \label{f3_asy_v} F_3 \sim 1.197 +
\frac{1.03}{p} - \frac{5.4}{p^2} +\frac{6.8}{p^3}-\frac{2.9}{p^4}.
\eeq We next observe that $F_3$  of the $(p,p)$ representations
behaves as $O(N)$. Thirdly we find that $F_3$ of the $(p,q)$
representations where $0< q <p$ behaves like that of $U(q+1)$ gauge
theory in the large $N$ limit when $q$ is fixed . It is because it
approach a constant which is consistent with the 2-loop
effective action of $U(q+1)$ gauge theory on $CP^2$: \beq (q+1)^3
F_3. \eeq

By assuming that we have correctly identified the large $N$ scaling
behavior of $F_3$ for various representations, we can obtain the
large $N$ limit of the effective actions after identifying the
suitable 't Hooft couplings for $CP^2$ and 6d manifolds. In the
$CP^2$ case, the action in the large $N$ limit is \beqa \Gamma &=& N
\left[ \frac{1}{2\lambda^2} + 6 \lambda^2 F_3 \right] , \n \lambda^2
&=& \frac{1}{f^4 N} . \eeqa In a 6d manifold of the $(p,p)$
representations, it is \beqa \Gamma &=& N^{\frac{4}{3}} \left[
\frac{3}{4\lambda^2} + 6 \lambda^2 \frac{F_3}{N} \right] , \n
\lambda^2 &=& \frac{1}{f^4 N^{\frac{1}{3}}} . \eeqa
Because of the
different large $N$ scaling behaviors of the effective actions, we
find that the $CP^2$ background is preferable to the 6d manifold.

After identifying the 't Hooft coupling, we can minimize the effective action
with respect to it.
We can use (\ref{ef_minac}) to determine the minimum of the effective action:
\beq
\Gamma \geq \Gamma_{min} \equiv 2 \sqrt{\Gamma_{tree} \Gamma_{2-loop}} .
\eeq

\begin{figure}[tbph]
\epsfysize=8cm
\begin{center}
\vspace{1cm}
\hspace{0cm}
\epsfbox{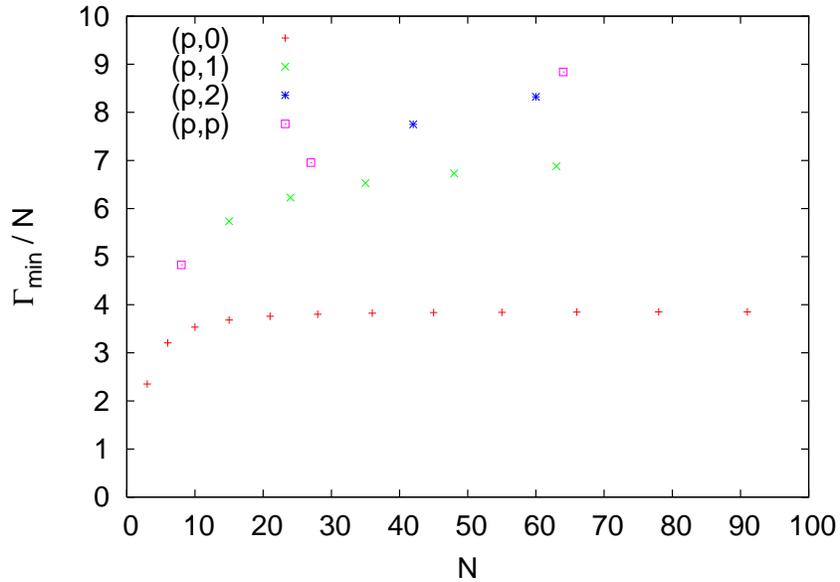}
\end{center}
\caption{$\Gamma_{min}/N$ against $N$.}
\label{fig:Eff}
\end{figure}

Fig.\ref{fig:Eff} shows $\Gamma_{min}/N$ against $N$. We can observe
that the effective action on the fuzzy $CP^2$ in the large $N$ limit is the
smallest in this class with $SU(3)$ symmetry as it approaches a
constant. This value can be estimated by using (\ref{f3_asy_v}) as \beq
\label{cp2_asym_val} \frac{\Gamma}{N} \simeq 3.79 \quad.
\label{effcp2} \eeq The 't Hooft coupling at this minimum is \beq
\lambda^2 \simeq 0.26 \ . \eeq

We remark here that (\ref{cp2_asym_val}) is comparable to the
minimum of the effective action of the fuzzy $S^2 \times S^2$
background at the most symmetric point \cite{fuzzy sphere}: \beq
\frac{\Gamma_{ S^2 \times S^2 }}{N} \simeq 3.61 \quad.
\label{effs2s2} \eeq Although we believe that the estimate
(\ref{effs2s2}) is accurate, our estimate (\ref{effcp2}) suffers
considerable uncertainty since it is derived from our numerical
investigation up to $N\sim 100$. As we observe in Table
\ref{ap_monf} that $F_{3}$ is gradually decreasing, we cannot
determine the lower bound of the effective action of $CP^2$ yet.
Within these limitations, we can still conclude that the fuzzy $CP^2$ background is
stable in its class and its effective action is comparable to that
of fuzzy $S^2 \times S^2$.

Here we summarize our findings for the backgrounds with $SU(3)$
symmetry. The effective action becomes $O(N)$ for the $(p,0)$
representations in the large $N$ limit. On the other hand the $(p,p)$
representations give the effective action $O(N^{\frac{3}{4}})$. We
recall here that the $(p,0)$ representations give a 4-dimensional NC
spacetime while the $(p,p)$ representations gives a 6-dimensional one
in the large $N$ limit. Since
the both effective actions are positive, the $(p,0)$ representations
are favored over the $(p,p)$ representations  in the large $N$ limit.
We also have an observation for the $(p,q)$ representations with
$q<<p$. In this case the $(p,q)$ representations behave like a
direct product of the $(p,0)$ representations and the
$(q+1)\times(q+1)$ identity matrix. In such a case, we effectively
obtain $U(q+1)$ gauge theory on $CP^2$ and the effective action is
proportional to $(q+1)^3N$. We thus argue that the effective action
is minimized for $q=0$. Therefore the $(p,0)$ representations are a
solution of I{}IB matrix model as long as $SU(3)$ symmetry is not
broken. We conclude that a four dimensional fuzzy $CP^2$ is
singled out by IIB matrix model within the manifolds with $SU(3)$
symmetry.

One of our goals of this paper is to investigate the scaling behavior of 
the effective action of this class of spacetime in the large $N$ limit. 
Let us recall the situation for the manifolds constructed from
$SU(2)$ algebras \cite{fuzzy sphere}.
The four dimensional fuzzy $S^2\times S^2$ makes
the effective action to be $O(N)$, and a six dimensional spacetime
$S^2\times S^2\times S^2$ gives $O(N^{\frac{4}{3}})$ action. These
scaling behaviors can be derived from the power counting of the higher loop
contributions.
We also assumed that the leading quantum corrections cancel
due to supersymmetry. Such an assumption can be justified since
the quantum corrections do cancel for commuting backgrounds
and the commutators of the backgrounds reduce the degrees of divergences.
In our identification of the 't Hooft couplings, we used the fact that the three
point vertices scales as $1/\sqrt{N}$ in the large $N$ limit. 

We argue that the same scaling rule holds in general. 
In fact our reasoning to identify the scaling behavior
of the effective action does not depend on the details
of a particular Lie algebra.
In particular, the
large $N$ scaling rule of the three point vertices are the
consequence of our normalization of the two point vertices to be
$O(1)$. Therefore it must hold in generic Lie algebra. 
In fact we have
numerically found, at the 2-loop level, that a 4-dimensional fuzzy $CP^2$, namely
the $(p,0)$ representation gives $O(N)$ effective action and a
6-dimensional fuzzy flag manifold, namely the $(p,p)$ representation
gives $O(N^{\frac{4}{3}})$ behavior. These findings
support our argument that any four dimensional fuzzy homogeneous
spacetime gives $O(N)$ effective action and six dimensional one gives
$O(N^{\frac{4}{3}})$ action.

We investigated whether I{}IB matrix model has a
fuzzy $S^2\times S^2$ solution at the 2-loop level previously. 
The most symmetric $S^2\times S^2$ solution turns out to be unstable along
some directions of their moduli parameters. They describe the
relative sizes of the two spheres. The instability drive the
symmetric $S^2\times S^2$ to the asymmetric one. Fortunately we find
fuzzy $CP^2$ has no such instability. The extremal value of the
effective action is comparable to that of the symmetric $S^2\times
S^2$. We thus obtain a new evidence for the existence of a symmetric
stable 4-dimensional spacetime in I{}IB matrix model.

\section{Conclusions} \label{conc}
In this paper we have investigated the effective action of IIB
matrix model on fuzzy $CP^2$ and the related manifold with $SU(3)$
isometry at the two loop level. Since the backgrounds constructed by
using $SU(3)$ algebra contain the manifolds with different
dimensionality such as $CP^2$ (4d) and a 6d manifold, we can compare
the minimum of the effective action of the 4d and 6d dimensional
backgrounds like \cite{fuzzy sphere} in our investigation of the
stability of $CP^2$.

We have investigated the large $N$ scaling behavior of the effective
action. The action scales as $N$ on $CP^2$ and $N^{\frac{4}{3}}$ on
a 6d manifold respectively. The effective action of the $(p,q)$
representations where $p > q$ with fixed  $q$ also scales as $N$,
since it behaves like $U(q+1)$ gauge theory of $CP^2$. From these
results, we have found that $CP^2$ minimizes the effective action
among the backgrounds which are constructed by $SU(3)$ algebra. We
conclude that the fuzzy $CP^2$ background is a solution in IIB
matrix model and stable as far as $SU(3)$ symmetry is not broken.

These scaling behaviors are in accord with other 4d manifolds like
$S^2\times S^2$ and $T^2\times T^2$ and also a 6d manifold
$S^2\times S^2\times S^2$\cite{fuzzy sphere,BHKK04}. These facts
support our contention that the effective action of a compact
manifold embedded in IIB matrix model has the universal scaling
behavior: it scales as $N$ and $N^{\frac{4}{3}}$ on a 4d and 6d
manifold respectively.

We have also compared the minimum of the effective actions of $CP^2$
with that of $S^2 \times S^2$. We have observed the effective action
of $CP^2$ is comparable to that of $S^2 \times S^2$. As we  have
observed in Table \ref{ap_monf} that the 2-loop effective action on
$CP^2$ is gradually decreasing, we cannot determine the lower bound
of it yet. Therefore we cannot say which is smaller even at the two
loop level. To answer this question, it is desirable to obtain 
an asymptotic expression of
the 2-loop effective action on $CP^2$ like such an expression on
$S^2$ which is obtained from the
Wigner's $6j$ symbols. Such an effort may be useful to determine
whether higher symmetry of the background may lower the effective
action or not.

\begin{center} \begin{large}
Acknowledgments
\end{large} \end{center}
This work is supported in part by the Grant-in-Aid for Scientific
Research from the Ministry of Education, Science and Culture of Japan.
We thank  T. Azuma, J. Nishimura and N. Kawahara for discussions.

\section*{Appendix A}
\renewcommand{\theequation}{A.\arabic{equation}}
\setcounter{equation}{0}

\subsubsection*{Construction of background}
A fundamental representation of SU(3) is 3-dimensional.
The Lie group generators can be written by Gell-Mann matrices $\lambda_{\mu}$ as
$t_{\mu}=\lambda_{\mu}/2$.
We take Gell-Mann matrices as the following form
\begin{eqnarray}
&&\lambda_1=\left(\begin{array}{ccc}
0&1&0\\
1&0&0\\
0&0&0
\end{array}\right),
\lambda_2=
\left(\begin{array}{ccc}
0&-i&0\\
i&0&0\\
0&0&0
\end{array}\right),
\lambda_3=
\left(\begin{array}{ccc}
1&0&0\\
0&-1&0\\
0&0&0
\end{array}\right), \n
&&\lambda_4=
\left(\begin{array}{ccc}
0&0&1\\
0&0&0\\
1&0&0
\end{array}\right),
\lambda_5=
\left(\begin{array}{ccc}
0&0&-i\\
0&0&0\\
i&0&0
\end{array}\right),
\lambda_6=
\left(\begin{array}{ccc}
0&0&0\\
0&0&1\\
0&1&0
\end{array}\right),\n
&&\lambda_7=
\left(\begin{array}{ccc}
0&0&0\\
0&0&-i\\
0&i&0
\end{array}\right),
\lambda_8=\frac{1}{\sqrt{3}}
\left(\begin{array}{ccc}
1&0&0\\
0&1&0\\
0&0&-2
\end{array}\right).
\end{eqnarray}

We denote state vectors on which these generators act as
$|a\rangle$,$|b\rangle$, ... ,
 here indices $a,b,...$ runs from $1$ to $3$.
These vectors have the following components
\begin{eqnarray}
|1\rangle\;=\;
\left(\begin{array}{c}
1\\0\\0
\end{array}\right),\quad
|2\rangle\;=\;
\left(\begin{array}{c}
0\\1\\0
\end{array}\right),\quad
|3\rangle\;=\;
\left(\begin{array}{c}
0\\0\\1
\end{array}\right).
\end{eqnarray}
The Cartan matrices are $t_3$ and $t_8$. They act on $|a\rangle$ as the following way
\begin{eqnarray}
&&t_3|1\rangle=\frac{1}{2}|1\rangle, \quad t_3|2\rangle=\frac{-1}{2}|2\rangle,
\quad t_3|3\rangle =0\cdot|3\rangle,\n
&&t_8|1\rangle=\frac{1}{2\sqrt{3}}|1\rangle,\quad t_8|2\rangle=\frac{1}{2\sqrt{3}}|2\rangle,\quad t_8|3\rangle=\frac{-1}{\sqrt{3}}|3\rangle.
\end{eqnarray}

The raising/lowering operators are
\begin{eqnarray}
j^{\pm}_1=t_4\pm it_5,\quad
j_{2}^{\pm}=t_6\pm it_7
\end{eqnarray}
and they act on the state vectors as
\begin{eqnarray}
j^{\pm}_{1}:
|3\rangle\;\leftrightarrow\; |1\rangle, \quad
j^{\pm}_{2}: |2\rangle\;\leftrightarrow\; |3\rangle, \quad
\mbox{otherwise gives zero.}
\end{eqnarray}

A general $SU(3)$ representation is labeled by a set of two integers $(p,q)$
and have the dimension
$dim(p,q)={(p+1)(q+1)(p+q+2)}/{2}$.
The fundamental representation is denoted as $(1,0)$.
The $(p,q)$ representation can be constructed from $(1,0)$ by forming tensor products.

As the first example, we construct the $(2,0)$ representation.
The $(2,0)$ state vectors are constructed from the tensor products of the two sets of 
the $(1,0)$ vectors:
\begin{eqnarray}
|v^{(2,0)}\rangle= |a\rangle|b\rangle+|b\rangle|a\rangle\;.
\end{eqnarray}
We should take an appropriate normalization factor in the above expression.
The symmetric property of this tensor product is represented by a Young tableau
\begin{picture}(25,10)
\put(0,-2){\framebox(20,10)}
\put(10,-2){\line(0,1){10}}
\end{picture}.
A single box
\begin{picture}(10,10)
\put(0,-2){\framebox(10,10)}
\end{picture}
denotes the $(1,0)$ vector.
The $(2,0)$ generators which act on the state vectors are the tensor products of
$(1,0)$ generators $t_{\mu}$ and the $3\times 3$ unit matrix ${\bf 1}_{3}$:
\begin{eqnarray}
T^{(2,0)}_{\mu}&=&t_{\mu}\otimes {\bf 1_{3}}+
{\bf 1_{3}}\otimes t_{\mu}\;.
\end{eqnarray}
To obtain the explicit matrix representation of the generators, 
we need to calculate the matrix elements
\begin{eqnarray}
\langle v^{(2,0)}|T^{(2,0)}|v^{(2,0)}\rangle\;.
\end{eqnarray}
In this way, we can write down the generators as $6\times 6$ matrices.
An extension to the $(p,0)$ representation is easily obtained by tensoring 
$p$ sets of the fundamental representations.
The $(p,0)$ state vectors up to the normalization factor are given by totally-symmetrized
tensor products of the $(1,0)$ vectors
\begin{eqnarray}
|v^{(p,0)}\rangle=
\prod^{p}_{i=1}|a_{i}\rangle
\;+\; \mbox{permutations for $\{a_{i}\}$}
\end{eqnarray}
Its symmetric property is represented by the Young tableau:
\begin{picture}(55,10)
\put(0,-2){\framebox(50,10)}
\put(10,-2){\line(0,1){10}}
\put(20,-2){\line(0,1){10}}
\put(40,-2){\line(0,1){10}}
\put(25,3){...}
\put(3,1){\tiny{1}}
\put(13,1){\tiny{2}}
\put(43,1){\tiny{p}}
\end{picture}.

The representations of the generators which act on these $(p,0)$ state vectors are
\begin{eqnarray}
T_{\mu}^{(p,0)}&=&
\sum_{i=0}^{p-1}\left({\bf 1}_3\otimes\right)^it_{\mu}
\left(\otimes {\bf 1}_3\right)^{p-1-i}
\end{eqnarray}
To obtain an explicit matrix representation of the generators, 
we need to calculate the matrix elements
\begin{eqnarray}
\langle v^{(p,0)}|T^{(p,0)}|v^{(p,0)}\rangle\;.
\end{eqnarray}
In this way, we can write down the generators as ${(p+1)(p+2)}/{2}\times
{(p+1)(p+2)}/{2}$ matrices.

Next we consider an extension of our construction to the $(p,p)$ representations.
It is obtained by $(2p+p)$-fold tensor products.
The state vectors of the $(p,p)$ representation up to the normalization factors can be written as
\begin{eqnarray}
|v^{(p,p)}\rangle&=&
\prod_{i=1}^{p}(\; |a_i\rangle |b_i \rangle - |b_i\rangle |a_i\rangle \;)
\prod^{p}_{j=1}|a_{j}\rangle
\;+\;\mbox{permutations of $\{a_i,a_j\}$}
\end{eqnarray}
Here the permutations between $a_i$ and $a_j$ also should be included.
Indices $a_i$ and $b_i$ are antisymmetrized.
Its symmetry property is represented by a Young tableau:
\begin{picture}(105,20)
\put(0,-5){\framebox(50,10)}
\put(0,5){\framebox(100,10)}
\put(10,-5){\line(0,1){20}}
\put(20,-5){\line(0,1){20}}
\put(40,-5){\line(0,1){20}}
\put(50,-5){\line(0,1){20}}
\put(60,5){\line(0,1){10}}
\put(70,5){\line(0,1){10}}
\put(90,5){\line(0,1){10}}
\put(3,8){\tiny{1}}
\put(13,8){\tiny{2}}
\put(25,8){...}
\put(43,8){\tiny{p}}
\put(75,8){...}
\put(92,8){\tiny{2p}}
\end{picture}.
The representations of the generators which act on these $(p,p)$ state vectors,
up to normalization factor, are
\begin{eqnarray}
T_{\mu}^{(p,p)}&=&
\sum_{i=0}^{3p-1}\left({\bf 1_3}\otimes \right)^it_{\mu}
\left(\otimes {\bf 1_3}\right)^{3p-1-i}
\end{eqnarray}
To obtain explicit form of the generators, we need to calculate the matrix elements
\begin{eqnarray}
\langle v^{(p,p)}|T^{(p,p)}|v^{(p,p)}\rangle\;.
\end{eqnarray}
In this way, we can write down the generators as $(p+1)^3\times (p+1)^3$ matrices.

An extension to an arbitrary $(p,q)$ representation is easily obtained
by forming the $(p+2q)$-fold tensor products.
The state vectors of $(p,q)$ type can be written as
\begin{eqnarray}
|v^{(p,q)}\rangle&=&
{\cal C}_{(p,q)}\prod_{i=1}^{q}
(\; |a_i\rangle |b_i \rangle - |b_i\rangle |a_i\rangle \;)
\prod^{p}_{j=1}|a_{j}\rangle
\;+\;\mbox{permutations of $\{a_i,a_j\}$}
\end{eqnarray}
Here the permutations between $a_i$ and $a_j$ should be included also.
Indices $a_i$ and $b_i$ are antisymmetrized.
The symmetric property is given by a Young tableau:
\begin{picture}(105,20)
\put(0,-5){\framebox(50,10)}
\put(0,5){\framebox(108,10)}
\put(10,-5){\line(0,1){20}}
\put(20,-5){\line(0,1){20}}
\put(40,-5){\line(0,1){20}}
\put(50,-5){\line(0,1){20}}
\put(60,5){\line(0,1){10}}
\put(70,5){\line(0,1){10}}
\put(90,5){\line(0,1){10}}
\put(3,8){\tiny{1}}
\put(13,8){\tiny{2}}
\put(25,8){...}
\put(43,8){\tiny{q}}
\put(75,8){...}
\put(91,8){\tiny{q+p}}
\end{picture}.
Here ${\cal C}_{(p,q)}$ is a normalization constant.
The representations of the generators which act on these $(p,q)$ state vectors are
\begin{eqnarray}
T_{\mu}^{(p,q)}&=&
\sum_{i=0}^{2p+q-1}\left( {\bf 1_3}\otimes\right)^it_{\mu}
\left(\otimes {\bf 1_3}\right)^{2p+q-1-i}
\end{eqnarray}
To obtain an explicit matrix form of the generators, we need to calculate the matrix elements
\begin{eqnarray}
\langle v^{(p,q)}|T^{(p,q)}|v^{(p,q)}\rangle\;.
\end{eqnarray}
In this way, we can write down the generators as $N^{(p,q)}\times N^{(p,q)}$
matrices
where
\beq
N^{(p,q)}=\frac{(p+1)(q+1)(p+q+2)}{2}.
\eeq

\subsubsection*{Construction of matrix harmonics in SU(3) background}
Suppose that we take a matrix model background to be a $(p,q)$ representation.
The gauge (and adjoint fermion) fields are expanded by harmonic matrices as follows
\begin{eqnarray}
\phi=\sum_{(A)}\sum_{ms}\;\phi_{\;ms}^{(A)}\;Y^{(A)}_{\;ms}\;,
\label{expansion}
\end{eqnarray}
where $Y^{(A)}_{ms}$ are the matrix harmonics.
The index $(A)$ denotes the sets of two integers $(p_A,q_A)$ which label the 
irreducible representations.
They are $ N^{(p,q)} \times N^{(p,q)}$ matrices which satisfy
\begin{eqnarray}
P_3 Y_{ms}^{(A)}&\equiv&[p_3, Y_{ms}^{(A)}]\;=\;m\;Y_{ms}^{(A)}\;,\n
P_8 Y_{ms}^{(A)}&\equiv&[p_8, Y_{ms}^{(A)}]\;=\;s\;Y_{ms}^{(A)}\;,\n
P^2 Y_{ms}^{(A)}&\equiv&[p_{\mu}[p_{\mu}, Y_{ms}^{(A)}]]\;=\;
(\frac{1}{2}p_A^2+p_A+\frac{1}{2}q_A^2+q_A)\;Y_{ms}^{(A)}\;.
\end{eqnarray}

The gauge fields are constructed as bi-local fields.
When the background is a $(p,q)$ representation, the bi-local state has a tensor
structure $(p,q)\otimes(q,p)$.
They can be decomposed into the irreducible representations, and
the decomposition may have the following form
\begin{eqnarray}
(p,q)\otimes (q,p)= \sum_{n=0}^{p+q}\;D_n\;(n,n)+
\sum_{l\neq m}^{p+2q}\;E_{ml}\;\big((l,m)+(m,l)\big)
\label{decomposition}
\end{eqnarray}
where $D_n$ and $E_{lm}$ are multiplicity factors.
If we take $q=0$, the decomposition becomes a simple form as
\begin{eqnarray}
{}(p,0)\otimes (0,p)&=&\sum_{n=0}^{p}(n,n)\;.
\end{eqnarray}
Here we give the $p=q=1$ case for another simple example
\begin{eqnarray}
{}(1,1)\otimes (1,1)&=&(2,2)+2(1,1)+(0,0)+(3,0)+(0,3)\;.
\end{eqnarray}
Thus, in expansion (\ref{expansion}), the sets of the integers $(p_A,q_A)$ run over 
the irreducible representations which appear in the decomposition,
and $m$ and $s$ take the value of these irreducible representations $(p_A,q_A)$.

Now we explain how to construct such matrices in a given background.
Let us describe a background (i.e. SU(3) generator of a $(p,q)$ rep.) in terms of
SU($N^{(p,q)}$) basis
\begin{eqnarray}
T_{\mu}^{(p,q)}&=& \sum_{\alpha}\;
({\cal A}_{\alpha}E_{\alpha}+{\cal B}_{-\alpha}E_{-\alpha})
+\sum_{a}{\cal C}_{a}H_{a}\;
\end{eqnarray}
where $\{E_{\alpha},E_{-\alpha},H_{\alpha}\}$ are Cartan's basis which satisfy
the following relations
\begin{eqnarray}
&&[H_a, H_b]=0\;,\n
&&{}[H_a,E_{\pm\alpha}]=\pm\alpha_{a}E_{\pm\alpha}\;,\n
&&{}[E_{\alpha},E_{-\alpha}]=\alpha^aH^{a},\quad
{}[E_{\alpha},E_{\beta}]=N_{\alpha,\beta}E_{\alpha+\beta}\;.
\qquad (E^{\dagger}_{\alpha}=E_{-\alpha})
\end{eqnarray}

One can take a representation of $T_{3}^{(p,q)}$ and $T_{8}^{(p,q)}$ as diagonal
matrices
\begin{eqnarray}
T^{(p,q)}_{3}=\sum_{a}{\cal C}_aH_a, \quad
T^{(p,q)}_{8}=\sum_{b}{\cal C}'_bH_b
\end{eqnarray}
Each $E_{\alpha}$s can be assigned to an off-diagonal matrix which has only one
non-zero component:
\begin{eqnarray}
(E_{\alpha})_{ij}=\Bigg\{
\begin{array}{cc}
1&\mbox{for} \quad(i,j)=(i_{\alpha},j_{\alpha}),\\
0&\mbox{otherwise}\;.
\end{array}
\end{eqnarray}
Then we have
\begin{eqnarray}
{}[T_{3}^{(p,q)},E_{\alpha}]= \sum_{a}{\cal C}_a \alpha^{a}E_{\alpha}, \quad
{}[T_{8}^{(p,q)},E_{\alpha}]=\sum_{b}{\cal C}'_b\alpha^{b}E_{\alpha}.
\end{eqnarray}
It implies that $Y^{(A)}_{\;ms}$ with $(m,s)=(\sum_{a}{\cal C}_a ,
\sum_{b}{\cal C}'_b )$ can be written as linear combinations of $E_{\alpha}$s
which have same eigenvalues of $(m,s)$.
On the other hand, Cartan subalgebra $[H,H]=0$ implies that $Y^{(A)}_{\;m=s=0}$
can be obtained by linear combinations of $H$.\\

Following the above observation,
we first take all commutators $[T^3,E]$ and $[T^8,E]$ to 
find quantum numbers $m$ and $s$ of each $E$'s.
Next we determine suitable linear combinations in the matrix basis 
which possess the same $m$ and $s$.
Then we obtain matrix harmonics which correspond to the irreducible representations 
in the decomposition (\ref{decomposition}).

One way to determine such linear combinations
is to use the raising/lowering operators.
 The decomposition (\ref{decomposition}) contains the irreducible representation $(p_A,q_A)=(p+2q,p-q)$.
The value $p+2q$ is the maximum value of $p_A$ in this decomposition.
The highest weight state is unique in each irreducible representation, and $p+2q$ is
the largest number in the decomposition.
Then there should be only one matrix base corresponding to such a state
whose eigenvalues are $m=\frac{1}{2}(p+2q+p-q)={2p+q}/{2}$ and
$s=\frac{1}{2\sqrt{3}}(p+2q+p-q-2(p-q))={3q}/{2\sqrt{3}}$.
Therefore a matrix base with the eigenvalues $m_0\equiv{2p+q}/{2}$ and
$s_0\equiv{3q}/{2\sqrt{3}}$
is uniquely identified with the highest weight state of $(p+2q,p-q)$.
Next we carry out the operations of the lowering operators and generate sets of
independent combinations of the matrix basis with
$m'( < m_0)$ and $s'(\neq s_0)$.
After suitable orthogonalizations, they form the state vectors with quantum number
$m'$ and $s'$.
Some of these belong to the $(p+2q,p-q)$ rep. and form $Y_{m's'}^{(p+2q,p-q)}$.
Others belong to different irreducible representations and form $Y_{m's'}^{(A')}$.
In this way, we can identify all $(A')\neq (p+2q,p-q)$ which appear in the decomposition (\ref{decomposition}).

There is another way to obtain suitable combinations of the matrix basis more straightforwardly.
First we collect matrix basis with the same quantum numbers $m$ and $s$ and
denote this set of basis as $\{w_i\}$.
Next we diagonalize the Casimir operator $P^2$ whose matrix elements are
\begin{eqnarray}
P^2_{ij}= tr(w_{i}^{\dagger} P^2w_j) \;.
\end{eqnarray}
A different eigenvalue of $P^2$ correspond to a different $(A)$ of $Y^{(A)}_{ms}$,
and $Y^{(A)}_{ms}$ themselves are obtained as the eigenvectors.
This method is useful if one has automatic computation tools for linear algebra,
like mathematica or maple etc.

\subsubsection*{An explicit example}
We give an explicit construction of a background (generators) and the matrix harmonics in
a simple case.
We consider the $(2,0)$ representation.

An expression of the state vectors of the $(2,0)$ representation is the following
\begin{eqnarray}
&&{}|1\rangle^{(2,0)}=|a\rangle|a\rangle\;,\quad
|2\rangle^{(2,0)}=\frac{|a\rangle|b\rangle+|b\rangle|a\rangle}{\sqrt{2}}\;,\quad
|3\rangle^{(2,0)}=|b\rangle|b\rangle\;,\n
&&{}|4\rangle^{(2,0)}= \frac{|a\rangle|c\rangle+|c\rangle|a\rangle}{\sqrt{2}}\;,\quad
|5\rangle^{(2,0)}=\frac{|b\rangle|c\rangle+|c\rangle|b\rangle}{\sqrt{2}}\;,\quad\n
&&{}|6\rangle^{(2,0)}=|c\rangle|c\rangle
\end{eqnarray}
where $ |a\rangle,|b\rangle$ and $|c\rangle$  are the state vectors of the fundamental representation.

The $SU(3)$ generators of the $(2,0)$ representation are
\begin{eqnarray}
&&T^{(2,0)}_3=
\left(\begin{array}{cccccc}
1&&&&&\\
&0&&&&\\
&&-1&&&\\
&&&\frac{1}{2}&&\\
&&&&-\frac{1}{2}&\\
&&&&&0\\
\end{array}\right),
T^{(2,0)}_8=
\left(\begin{array}{cccccc}
\frac{1}{\sqrt{3}}&&&&&\\
&\frac{1}{\sqrt{3}}&&&&\\

&&\frac{1}{\sqrt{3}}&&&\\
&&&-\frac{1}{2\sqrt{3}}&&\\
&&&&-\frac{1}{2\sqrt{3}}&\\
&&&&&-\frac{2}{\sqrt{3}}\\
\end{array}\right),\n
\n
&&T^{(2,0)}_1=
\left(\begin{array}{cccccc}
0&\frac{1}{\sqrt{2}}&0&0&0&0\\
\frac{1}{\sqrt{2}}&0&\frac{1}{\sqrt{2}}&0&0&0\\
0&\frac{1}{\sqrt{2}}&0&0&0&0\\
0&0&0&0&\frac{1}{2}&0\\
0&0&0&\frac{1}{2}&0&0\\
0&0&0&0&0&0\\
\end{array}\right),
T^{(2,0)}_2=
\left(\begin{array}{cccccc}
0&\frac{-i}{\sqrt{2}}&0&0&0&0\\
\frac{i}{\sqrt{2}}&0&\frac{-i}{\sqrt{2}}&0&0&0\\
0&\frac{i}{\sqrt{2}}&0&0&0&0\\
0&0&0&0&\frac{-i}{2}&0\\
0&0&0&\frac{i}{2}&0&0\\
0&0&0&0&0&0\\
\end{array}\right),\n
&&T^{(2,0)}_4=
\left(\begin{array}{cccccc}
0&0&0&\frac{1}{\sqrt{2}}&0&0\\
0&0&0&0&\frac{1}{2}&0\\
0&0&0&0&0&0\\
\frac{1}{\sqrt{2}}&0&0&0&0&\frac{1}{\sqrt{2}}\\
0&\frac{1}{2}&0&0&0&0\\
0&0&0&\frac{1}{\sqrt{2}}&0&0
\end{array}\right),
T^{(2,0)}_5=
\left(\begin{array}{cccccc}
0&0&0&\frac{-i}{\sqrt{2}}&0&0\\
0&0&0&0&\frac{-i}{2}&0\\
0&0&0&0&0&0\\
\frac{i}{\sqrt{2}}&0&0&0&0&\frac{-i}{\sqrt{2}}\\
0&\frac{i}{2}&0&0&0&0\\
0&0&0&\frac{i}{\sqrt{2}}&0&0\\
\end{array}\right),\n
&&T^{(2,0)}_6=\left(\begin{array}{cccccc}
0&0&0&0&0&0\\
0&0&0&\frac{1}{2}&0&0\\
0&0&0&0&\frac{1}{\sqrt{2}}&0\\
0&\frac{1}{2}&0&0&0&0\\

0&0&\frac{1}{\sqrt{2}}&0&0&\frac{1}{\sqrt{2}}\\
0&0&0&0&\frac{1}{\sqrt{2}}&0\\
\end{array}\right),
T^{(2,0)}_7=\left(\begin{array}{cccccc}
0&0&0&0&0&0\\
0&0&0&\frac{-i}{2}&0&0\\
0&0&0&0&\frac{-i}{\sqrt{2}}&0\\
0&\frac{i}{2}&0&0&0&0\\
0&0&\frac{i}{\sqrt{2}}&0&0&\frac{-i}{\sqrt{2}}\\
0&0&0&0&\frac{i}{\sqrt{2}}&0\\
\end{array}\right).
\end{eqnarray}

To construct matrix harmonics,
we define  off-diagonal matrix bases as
\begin{eqnarray}
\left(\begin{array}{cccccc}
0&E_{\alpha_1}&E_{\alpha_2}&E_{\alpha_3}&E_{\alpha_4}&E_{\alpha_5}\\
E_{-\alpha_1}&0&E_{\alpha_6}&E_{\alpha_7}&E_{\alpha_8}&E_{\alpha_9}\\
E_{-\alpha_{2}}&E_{-\alpha_6}&0&E_{\alpha_{10}}&E_{\alpha_{11}}&E_{\alpha_{12}}\\
E_{-\alpha_3}&E_{-\alpha_7}&E_{-\alpha_{10}}&0&E_{\alpha_{13}}&E_{\alpha_{14}}\\
E_{-\alpha_4}&E_{-\alpha_8}&E_{-\alpha_{11}}&E_{-\alpha_{13}}&0&E_{\alpha_{15}}\\
E_{-\alpha_5}&E_{-\alpha_9}&E_{-\alpha_{12}}&E_{-\alpha_{14}}&E_{-\alpha_{15}}&0
\end{array}\right)\;.
\end{eqnarray}
This notation means that $E_{\alpha_1}$ is given by the form:
\begin{eqnarray}
E_{\alpha_1}=
\left(\begin{array}{cccccc}
0&1&0&0&0&0 \\
0&0&0&0&0&0 \\
0&0&0&0&0&0 \\
0&0&0&0&0&0 \\
0&0&0&0&0&0 \\
0&0&0&0&0&0
\end{array}\right)
\end{eqnarray}
and so on.

Following the decomposition
\begin{eqnarray}
(2,0)\otimes (0,2)=(2,2)+(1,1)+(0,0) \;,
\end{eqnarray}
we construct $Y^{(2,2)}$, $Y^{(1,1)}$ and $Y^{(0,0)}$ using the above matrix bases
and diagonal matrices.

Here is a Result. $(2,2)$ is $27$ dimensional:
\begin{eqnarray}
&&
Y^{(2,2)}_{1,\sqrt{3}}= E_{\alpha_{5}}\;,\;
Y^{(2,2)}_{0,\sqrt{3}}= E_{\alpha_{9}}\;,\;
Y^{(2,2)}_{-1,\sqrt{3}}= E_{\alpha_{12}}\;,\n
\n
&&
Y^{(2,2)}_{\frac{3}{2}, \frac{\sqrt{3}}{2} }= -E_{\alpha_{4}}\;,\;
Y^{(2,2)}_{\frac{1}{2},\frac{\sqrt{3}}{2} }=
\left\{\begin{array}{c}
\frac{-E_{\alpha_{3}}+E_{\alpha_{14}}}{\sqrt{2}}\\
\frac{ E_{\alpha_{3}} +E_{\alpha_{14}} -2\sqrt{2} E_{\alpha_{8}}}{\sqrt{10}}
\end{array}\right.\;,\;
Y^{(2,2)}_{\frac{-1}{2},\frac{\sqrt{3}}{2} }=
\left\{\begin{array}{c}
\frac{-E_{\alpha_{11}}+E_{\alpha_{15}}}{\sqrt{2}}\\
\frac{
-E_{\alpha_{11}}-E_{\alpha_{15}}+2\sqrt{2}E_{\alpha_{7}}
}{\sqrt{10}}
\end{array}\right.\;,\;\n
&&
Y^{(2,2)}_{\frac{-3}{2},\frac{\sqrt{3}}{2} }= -E_{\alpha_{10}}\;,\;\n
\n
&&Y^{(2,2)}_{2,0}=E_{\alpha_2}\;,\;
Y^{(2,2)}_{1,0}=\left\{
\begin{array}{c}
\frac{E_{\alpha_1}-\sqrt{2}E_{\alpha_{13}}}{\sqrt{3}}\\
\frac{-2E_{\alpha_1}-\sqrt{2}E_{\alpha_{13}}+3E_{\alpha_{6}}}{\sqrt{15}}
\end{array}\right.\;,\;
Y^{(2,2)}_{0,0}=\left\{
\begin{array}{c}
diag(0,0,1,0,-2,1)/\sqrt{6}\\
diag(0,2,-1,-2,0,1)/\sqrt{10}\\
diag(3,-3,1,-3,1,1)/\sqrt{30}\\
\end{array}\right. \;,\n
&&Y^{(2,2)}_{-1,0}=\left\{
\begin{array}{c}
\frac{E_{-\alpha_1}-\sqrt{2}E_{-\alpha_{13}}}{\sqrt{3}}\\
\frac{-2E_{-\alpha_1}-\sqrt{2}E_{-\alpha_{13}}+3E_{-\alpha_{6}}}{\sqrt{15}}
\end{array}\right.\;,\;
Y^{(2,2)}_{-2,0}=E_{-\alpha_2}\;,\;\n
\n
&&
Y^{(2,2)}_{\frac{3}{2},\frac{-\sqrt{3}}{2} }= E_{-\alpha_{10}}\;,\;
Y^{(2,2)}_{\frac{1}{2},\frac{-\sqrt{3}}{2} }=
\left\{\begin{array}{c}
\frac{E_{-\alpha_{11}}-E_{-\alpha_{15}}}{\sqrt{2}}\\
\frac{
-E_{-\alpha_{11}}-E_{-\alpha_{15}}+2\sqrt{2}E_{-\alpha_{7}}
}{\sqrt{10}}
\end{array}\right.\;,\;\n
\n
&&Y^{(2,2)}_{\frac{-1}{2},\frac{-\sqrt{3}}{2} }=
\left\{\begin{array}{c}
\frac{E_{-\alpha_{3}}-E_{-\alpha_{14}}}{\sqrt{2}}\\
\frac{ -E_{-\alpha_{3}} -E_{-\alpha_{14}} +2\sqrt{2} E_{-\alpha_{8}}}{\sqrt{10}}
\end{array}\right.\;,\;
Y^{(2,2)}_{\frac{-3}{2}, \frac{-\sqrt{3}}{2} }= E_{-\alpha_{4}}\;,\;\n
\n
&&
Y^{(2,2)}_{1,-\sqrt{3}}= E_{-\alpha_{12}}\;,\;
Y^{(2,2)}_{0,-\sqrt{3}}= E_{-\alpha_{9}}\;,\;
Y^{(2,2)}_{-1,-\sqrt{3}}= E_{-\alpha_{5}}
\end{eqnarray} .
$(1,1)$ is $8$ dimensional:
\begin{eqnarray}
&&
Y^{(1,1)}_{\frac{1}{2},\frac{\sqrt{3}}{2}}=
-\frac{\sqrt{2}E_{\alpha_{3}}+\sqrt{2}E_{\alpha_{14}}+E_{\alpha_{8}} }{\sqrt{5}}
\;,\;
Y^{(1,1)}_{\frac{-1}{2},\frac{\sqrt{3}}{2}}=
-\frac{ E_{\alpha_{7}}+\sqrt{2}E_{\alpha_{11}}+\sqrt{2}E_{\alpha_{15}} }{\sqrt{5}}
\;,\;\n
\n
&&
Y^{(1,1)}_{1,0}= \frac{ 2E_{\alpha_{1}}+\sqrt{2}E_{\alpha_{13}} +2E_{\alpha_{6}}  }
{\sqrt{10}}\;,\;
Y^{(1,1)}_{(0,0)}=
\left\{\begin{array}{c}
\frac{diag(0,1,2,-1,0,-2)}{\sqrt{10}}\\
\frac{diag(4,1,-2,1,-2,-2)}{\sqrt{30}}
\end{array}\right. \;,\;\n
&&
Y^{(1,1)}_{-1,0}= \frac{ 2E_{-\alpha_{1}}+\sqrt{2}E_{-\alpha_{13}} +
2E_{-\alpha_{6}}  }{\sqrt{10}}\;,\;\n
\n
&&
Y^{(1,1)}_{\frac{1}{2},\frac{-\sqrt{3}}{2}}=
\frac{ E_{-\alpha_{7}}+\sqrt{2}E_{-\alpha_{11}}+\sqrt{2}E_{-\alpha_{15}} }{\sqrt{5}}
\;,\;
Y^{(1,1)}_{\frac{-1}{2},\frac{-\sqrt{3}}{2}}=
\frac{\sqrt{2}E_{-\alpha_{3}}+\sqrt{2}E_{-\alpha_{14}}+E_{-\alpha_{8}} }{\sqrt{5}} .\n
\end{eqnarray}
Finally there is the singlet correspond to (0,0) :
\begin{eqnarray}
Y^{(0,0)}_{(0,0)}=\frac{1}{\sqrt{6}}{\bf 1}_{6}\;.
\end{eqnarray}

The 2-loop contribution to the effective action 
is calculated with these harmonics. The planar contribution is
\begin{eqnarray}
6n^3\sum_{(n_1,n_2,n_3)=1}^{2}\sum_{m_1,s_1}\sum_{m_2,s_2}\sum_{m_3,s_3}\;
\frac{ tr(Y^{(n_1,n_1)}_{m_1,s_1}Y^{(n_2,n_2)}_{m_2,s_2}Y^{(n_3,n_3)}_{m_3,s_3})
tr(Y^{(n_3,n_3)\dagger}_{m_3,s_3}Y^{(n_2,n_2)\dagger}_{m_2,s_2}
Y^{(n_1,n_1)\dagger}_{m_1,s_1})
 }{ n_1(n_1+1)n_2(n_2+1) n_3(n_3+1)  } \n
\end{eqnarray}
in $U(n)$ gauge theory.
On the other hand, the non-planar contribution is
\begin{eqnarray}
-6n\sum_{(n_1,n_2,n_3)=1}^{2}\sum_{m_1,s_1}\sum_{m_2,s_2}\sum_{m_3,s_3}\;
\frac{ tr(Y^{(n_1,n_1)}_{m_1,s_1}Y^{(n_2,n_2)}_{m_2,s_2}Y^{(n_3,n_3)}_{m_3,s_3})
tr(Y^{(n_1,n_1)\dagger}_{m_1,s_1}Y^{(n_2,n_2)\dagger}_{m_2,s_2}
Y^{(n_3,n_3)\dagger}_{m_3,s_3})
 }{ n_1(n_1+1)n_2(n_2+1) n_3(n_3+1)  }\;. \n
\end{eqnarray}

By substituting the explicit form of $Y^{(n,n)}_{ms}$, we obtain the planar contribution:
\begin{eqnarray}
6n^3\;\frac{42605}{41472}
\end{eqnarray}
and the non-planar contribution:
\begin{eqnarray}
-6n\;\frac{1115}{41472}\;.
\end{eqnarray}

\section*{Appendix B}
\renewcommand{\theequation}{B.\arabic{equation}}
\setcounter{equation}{0}
In this appendix, we evaluate the two loop effective action of IIB matrix model
in a
 fuzzy background which is made from a $(p,q)$ rep. of the $SU(3)$ generators
(\ref{eff-2loop}).

In this calculation, we make use of the following relation:
\beq \label{compPY}
\sum_m P_{\mu} Y^{(r,s)}_{m}{}^{\dagger} P_{\nu} Y^{(r,s)}_{m} = - \sum_m
Y^{(r,s)}_{m}{}^{\dagger} P_{\nu} P_{\mu} Y^{(r,s)}_{m},
\eeq
where the subscript $(r,s)$ denotes an irreducible representation of $SU(3)$ and $m$ 
denotes the eigenvalues of the Cartan subalgebra in the $(r,s)$ representation.
We first note that the harmonic matrices of $SU(3)$ obey the orthogonal relations:
\beq \label{orth-HM}
Tr \left( Y^{(r,s)}_{m}{}^{\dagger} Y^{(r',s')}_{m'} \right) =
\delta_{(r,s),(r',s')} \delta_{m,m'} .
\eeq
Let us perform a unitary transformation on $Y^{(r,s)}_{m}$:
\beqa \label{unitraY}
Y^{(r,s)}_{m} &\rightarrow& U Y^{(r,s)}_{m} U^{\dagger} = \sum_{n} u_{mn}
Y^{(r,s)}_{n} , \n
Y^{(r,s)}_{m}{}^{\dagger} &\rightarrow& \left( U Y^{(r,s)}_{m} U^{\dagger}
\right)^{\dagger}
 = \sum_{n} u^{*}_{mn} Y^{(r,s)}_{n}{}^{\dagger},
\eeqa
where $U$ is an $N \times N$ unitary matrix and $u_{mn}$ is the unitary
transformation represented in the $m$ bases.
Under (\ref{unitraY}), (\ref{orth-HM}) is transformed as
\beqa
Tr \left( Y^{(r,s)}_{m}{}^{\dagger} Y^{(r',s')}_{m'} \right)
&\rightarrow& \sum_{n,n'} u^{*}_{mn} u_{m'n'} Tr \left(
Y^{(r,s)}_{n}{}^{\dagger} Y^{(r',s')}_{n'} \right) \n
&& = \sum_{n,n'} u^{*}_{mn} u_{m'n'} \delta_{nn'} = \left( u u^{\dagger}
\right)_{m'm} \delta_{(r,s),(r',s')} .
\eeqa
Since (\ref{orth-HM}) is apparently invariant under
(\ref{unitraY}),
we can obtain
\beq \label{orthuu}
\left( u u^{\dagger} \right)_{m'm} = \delta_{m'm} .
\eeq
Using this relation, we can show that $\sum_{m} Y^{(r,s)}_{m}{}^{\dagger} Y^{(r,s)}_{m}$
is invariant under (\ref{unitraY}):
\beqa \label{compY}
\sum_{m} Y^{(r,s)}_{m}{}^{\dagger} Y^{(r,s)}_{m}
&\rightarrow& \sum_{m,n,n'} u^{*}_{mn} u_{mn'} Y^{(r,s)}_{n}{}^{\dagger}
Y^{(r,s)}_{n'} \n
&&= \sum_{n} Y^{(r,s)}_{n}{}^{\dagger} Y^{(r,s)}_{n}
\eeqa
Since $P_{\mu}$ are the generators of $SU(3)$ transformation, (\ref{compY}) is
equivalent to
\beq
P_{\mu} \left( \sum_{m} Y^{(r,s)}_{m}{}^{\dagger} Y^{(r,s)}_{m} \right) =0.
\eeq
From this formula, we can obtain (\ref{compPY}).

We introduce the wave functions and averages as
\beqa
\Psi_{123} &\equiv& Tr (Y^{(r_1,s_1)}_{m_1}Y^{(r_2,s_2)}_{m_2}Y^{(r_3,s_3)}_{m_3}) , \n
\left< X \right>_P &\equiv& \sum_{(r_i,s_i),m_i} \Psi_{123}^{*} X \Psi_{123} , \n
P_{i}^{\mu} Y^{(r_1,s_1)}_{m_1} &\equiv& [p_{\mu},Y^{(r_1,s_1)}_{m_1}] .
\eeqa
Where the sum of $(r_i,s_i)$ runs over the representations which are made from the
product of $(p,q)$ and $(q,p)$.
We introduce the following quantity
\beq
f_{\mu \nu \rho} f_{\mu \nu \sigma} = C_{G} \delta_{\rho \sigma},
\eeq
where $C_G$ is a constant which assumes $C_G=2$ for $SU(2)$ and $C_G=3$ for $SU(3)$.
With these preparations, we can calculate the
two loop effective action almost the same way as the fuzzy sphere case.

We expand quantum fluctuations in terms of the harmonic matrices:
\beqa
\mbox{gauge boson}&&
a^{\mu} \;=\; \sum_{(r,s),m} a^{(r,s) \ \mu}_{m} Y^{(r,s)}_{m} , \n
\mbox{fermion}&&
\varphi \;=\;\sum_{(r,s),m} \varphi^{(r,s)}_{m} Y^{(r,s)}_{m} , \n
\mbox{anti-ghost}&&
b \;=\; \sum_{(r,s),m} b^{(r,s)}_{m} Y^{(r,s)}_{m} , \n
\mbox{ghost}&&
c \;=\; \sum_{(r,s),m} c^{(r,s)}_{m} Y^{(r,s)}_{m} .
\eeqa
Then the propagators are derived from the kinematic terms:
\beqa
\left< a^{\mu} a^{\nu} \right> &=& \sum_{(r,s),m}
\left( P^2 \delta_{\mu \nu} + 2i f_{\mu \nu \rho} P^{\rho} \right)^{-1}
Y^{(r,s)}_{m} \ Y^{(r,s)}_{m}{}^{\dagger} , \n
\left< \varphi \bar{\varphi} \right> &=&  \sum_{(r,s),m}
\left( - \Gamma_{\mu} P_{\mu} \right)^{-1}
Y^{(r,s)}_{m} \ Y^{(r,s)}_{m}{}^{\dagger} , \n
<cb> &=& \sum_{(r,s),m} \frac{1}{P^2}
Y^{(r,s)}_{m} \ Y^{(r,s)}_{m}{}^{\dagger} .
\eeqa
We exclude the singlet state $(0,0)$ in the propagator.
To calculate the leading contributions in the large $N$ limit,
we expand the boson and the fermion propagators as
\beqa
\left( P^2 \delta_{\mu \nu} + 2i f_{\mu \nu \rho} P^{\rho} \right)^{-1} &\simeq&
\frac{\delta_{\mu \nu}}{P^2} - 2i \frac{f_{\mu \nu \rho} P^{\rho}}{P^4}
+ 4 \frac{I_{\mu \nu}(P)}{P^6} , \n
\left( - \Gamma_{\mu} P_{\mu} \right)^{-1} &\simeq&
\frac{\Gamma^{\mu} P_{\mu} }{P^2} +
\frac{i}{2}\frac{f_{\mu \nu \sigma}\Gamma^{\mu \nu \rho} P_{\sigma} P_{\rho} }{P^4} .
\eeqa
We have introduced the following tensor
\beq
I_{\mu \nu} \equiv f_{\tau \mu \rho}f_{\tau \nu \sigma} P_{\rho}P_{\sigma}.
\eeq
Using these propagators, we can calculate the contributions to the two loop effective action
from various interaction vertices as follows.

4-gauge boson vertex is
\begin{equation}
V_4 = - \frac{1}{4} Tr [a_\mu ,a_\nu ]^2 .
\end{equation}
The leading contribution to the two loop effective action is
\beq \label{ap_4b}
<-V_4> = -45F_1 -42 C_G G_1 + 3 C_G G_2 .
\eeq
Here
\beqa
F_1 &=& \left< \frac{1}{P_1^4 P_2^4} \right>_P , \n
G_1 &=& \left< \frac{1}{P_1^4 P_2^2 } \right>_P , \n
G_2 &=& \left< \frac{P_3^2}{P_1^4 P_2^4} \right>_P .
\eeqa

Ghost vertex is
\begin{equation}
V_g = Tr b \left[ p_{\mu} , \left[  a_{\mu} ,c \right] \right] .
\end{equation}
Their contribution is
\begin{equation} \label{ap_gh}
\frac{1}{2} < V_g V_g > = F_2 + 4 H_2 .
\end{equation}
Here
\beqa
F_2 &=& \left< \frac{P_2 \cdot P_3}{P_1^2 P_2^2 P_3^2} \right>_P , \n
H_2 &=& \left< \frac{P_2 \cdot I(1) \cdot P_3}{P_1^6 P_2^2 P_3^2} \right>_P ,
\eeqa
and
\begin{equation}
P_i \cdot I(j) \cdot P_k \equiv P_i^{\mu} I_{\mu \nu} (P_j) P_k^{\nu} .
\end{equation}

3-gauge boson vertex is
\begin{equation}
V_3 = - Tr P_{\mu} a_{\nu} \left[ a_{\mu}, a_{\nu} \right] .
\end{equation}
Their contribution is
\beqa
\label{ap_3b}
\frac{1}{2} < V_3 V_3 > &=& 9F_1 -9F_2 + C_G(6F_3 +2G_1+G_2) \n
&&+32H_1 - 36 H_2 -16H_3 +12H_4 -4H_5.
\eeqa
Newly introduced functions are defined as
\beqa
F_3 &=& \left< \frac{1}{P_1^2 P_2^2 P_3^2} \right>_P , \n
G'_1 &=& G_1-\frac{1}{N} Tr\left[ \left( \frac{1}{P^2} \right)^3 \right] , \n
H_1 &=& \left< \frac{P_1 \cdot I(2) \cdot P_1}{P_1^2 P_2^6 P_3^2} \right>_P , \n
H_3 &=& \left< \frac{P_2 \cdot I(1) \cdot P_3}{P_1^4 P_2^4 P_3^2} \right>_P , \n
H_4 &=& \left< \frac{P_1 \cdot I(2) \cdot P_1}{P_1^4 P_2^4 P_3^2} \right>_P , \n
H_5 &=& \left< \frac{P_2 \cdot I(1) \cdot P_3}{P_1^2 P_2^4 P_3^4} \right>_P .
\eeqa
In SU(3), we can evaluate the following quantity as
\beq
\frac{1}{N} Tr\left[ \left( \frac{1}{P^2} \right)^3 \right] = \frac{1}{N}
\sum_{(r,s),m} \frac{\frac{1}{2} (r+1)(s+1)(r+s+2)}{ \left[ \frac{1}{2} (r(r+2)+s(s+2))
\right]^3 } .
\eeq

Fermion vertex is
\begin{equation}
V_f=- \frac{1}{2} Tr \bar{\varphi} \Gamma_\mu \left[ a_\mu , \varphi \right] .
\end{equation}
Their contribution is
\beqa
\label{ap_fr}
\frac{1}{2} < V_f V_f > &=& -64F_2 +(-8C_G G'_1 +4C_GG_2 +8C_GF_3 +32H_4 ) \n
&&-16C_GF_3 +48C_GG'_1 + -8C_G G_2 +64 H_2 +64H_3 .
\eeqa

After summing up (\ref{ap_4b}), (\ref{ap_gh}), (\ref{ap_3b}) and (\ref{ap_fr}),
we find the 2-loop effective action:
\beqa  \label{ap_res}
\Gamma_{2-loop} &=& 2 C_G F_3 + 32H_1 +32H_2+48H_3+(12+32)H_4 -4H_5 \n
&=& 2 C_G F_3 .
\eeqa
It is because
\beqa
H_1+H_2 &=& 0 , \n
H_3+H_4 &=& 0 , \n
H_3-H_5 &=& 0 .
\eeqa

Since we have used the common properties of $SU(2)$ and $SU(3)$,
 the result (\ref{ap_res}) is valid for $SU(2)$ and $SU(3)$
 and consistent with the fuzzy sphere's results.

\section*{Appendix C}
\renewcommand{\theequation}{C.\arabic{equation}}
\setcounter{equation}{0}

In this appendix, we calculate $F_3$ in (\ref{ap_res}) numerically.
A practical way to calculate $F_3$ is to use Monte-Carlo simulation \cite{MonteCalro}.
Our strategy is to construct a Gaussian matrix model to calculate it:
\beqa
F_3 &=& \left< \frac{1}{P_1^2 P_2^2 P_3^2} \right>_P \n
&=& \int da db dc Tr(abc) Tr(cba) \exp \left( -\frac{1}{2} \left[ a,p^{\mu} \right]^2
-\frac{1}{2} \left[ b,p^{\mu} \right]^2 -\frac{1}{2} \left[ c,p^{\mu} \right]^2\right).
\eeqa
We can use the heat-bath algorithm to calculate this correlator.
The result is shown in Table \ref{ap_monf}.
We estimate the statical errors using a jackknife method \cite{JackkNife,MonteCalro}.

\begin{table}[tbph]
\caption{The results of $F_3$ using Monte-Calro simulation.} \label{ap_monf}
\begin{minipage}{8cm}
\begin{tabular}{|c|c|c|}

\hline
$SU(3)$ rep. & $N$ & $F_3$ \\
\hline
(1,0) & 3  & 0.69152 +/- 0.00056 \\
(2,0) & 6  & 1.02763 +/- 0.00064 \\
(3,0) & 10 & 1.15620 +/- 0.00069 \\
(4,0) & 15 & 1.21168 +/- 0.00072 \\
(5,0) & 21 & 1.23653 +/- 0.00072 \\
(6,0) & 28 & 1.24858 +/- 0.00071 \\
(7,0) & 36 & 1.25357 +/- 0.00073 \\
(8,0) & 45 & 1.25474 +/- 0.00086 \\
(9,0) & 55 & 1.25222 +/- 0.00091 \\
(10,0) & 66 & 1.25201 +/- 0.00088 \\
(11,0) & 78 & 1.25188 +/- 0.00091 \\
(12,0) & 91 & 1.24959 +/- 0.00091 \\
\hline
\end{tabular}
\end{minipage}
\hfill
\begin{minipage}{8cm}
\begin{tabular}{|c|c|c|}
\hline
$SU(3)$ rep. & $N$ & $F_3$ \\
\hline
(1,1) & 8  & 3.4551 +/- 0.0020 \\
(2,1) & 15 & 5.1412  +/- 0.0031 \\
(3,1) & 24 & 6.2030  +/- 0.0043 \\
(4,1) & 35 & 6.9072  +/- 0.0048 \\
(5,1) & 48 & 7.3973  +/- 0.0051 \\
(6,1) & 63 & 7.7632  +/- 0.0054 \\
(2,2) & 27 & 9.0688  +/- 0.0051 \\
(3,2) & 42 & 12.366  +/- 0.0086 \\
(4,2) & 60 & 15.064  +/- 0.011 \\
(3,3) & 64 & 18.522  +/- 0.013 \\
\hline
\end{tabular}
\end{minipage}
\end{table}

The other way to calculate $F_3$ is to use the harmonic matrices.
We can obtain these matrices on the computer using the method explained in appendix A.
The result is shown in Table \ref{ap_eigm}.

\begin{table}[tbph]
\caption{The results of $F_3$ using the harmonic matrices.} \label{ap_eigm}
\begin{center}
\begin{tabular}{|c|c|c|}
\hline
$SU(3)$ rep. & $N$ & $F_3$ \\
\hline
(1,0) & 3  & 0.691358 \\
(2,0) & 6  & 1.027320 \\
(3,0) & 10 & 1.156321 \\
(4,0) & 15 & 1.211689 \\
(5,0) & 21 & 1.236921 \\
(6,0) & 28 & 1.248420 \\
\hline
\end{tabular}
\end{center}
\end{table}

Since Table \ref{ap_eigm} shows the exact results, this calculation is preferable to the
Monte-Carlo.
But we have used the Monte-Carlo method,
because the exact evaluation requires more computer power than the Monte-Carlo.
Nevertheless we can use Table \ref{ap_eigm} to check Table \ref{ap_monf}.
We can thus claim that the Monte-Carlo method gives the correct results.

\end{document}